\documentclass[9pt,technote]{IEEEtran}

\usepackage{graphicx}
\usepackage{times}
\usepackage{epsfig}
\usepackage{amsmath}
\usepackage{amssymb}
\usepackage{algorithm}
\usepackage{algorithmic}
\usepackage{color}
\usepackage{cite}

\newcommand{\eq}{\triangleq}

\newcommand{\hfs}{\hfill\ensuremath{\square}} 
\DeclareMathOperator*{\argmin}{arg\,min}
\DeclareMathOperator*{\supp}{supp}

\newcommand{\field}[1]{\mathbb{#1}}
\newcommand{\R}{\field{R}}
\newcommand{\N}{\field{N}}

\newcommand{\F}{{\mathcal{U}}}
\newcommand{\E}{{\mathcal{E}}}

\newcommand{\T}{\top}

\newcommand{\vc}[1]{{\boldsymbol{#1}}}
\newcommand{\K}{K}

\newcommand{\Bs}{B}

\newcommand{\vx}{\vc{x}}
\newcommand{\vu}{\vc{u}}
\newcommand{\vz}{\vc{0}}
\newcommand{\vv}{\vc{v}}
\newcommand{\vf}{\vc{f}}
\newcommand{\eps}{\vc{\varepsilon}}

\newcommand{\elll}{\ell^1\text{-}\ell^2}

\newtheorem{thm}{Theorem}

\newtheorem{lem}[thm]{Lemma}
\newtheorem{prop}[thm]{Proposition}

\newtheorem{rem}[thm]{Remark}
\newtheorem{ass}[thm]{Assumption}

\begin{document}

 \title{Sparse Packetized Predictive Control for
 Networked Control over Erasure Channels}

\author{Masaaki Nagahara,~\IEEEmembership{Member,~IEEE,}\\
	Daniel E.~Quevedo,~\IEEEmembership{Member,~IEEE,}\\
	Jan {\O}stergaard,~\IEEEmembership{Senior Member,~IEEE}%
\thanks{M.\ Nagahara is with
	Graduate School of Informatics,
	Kyoto University, Kyoto, 606-8501, Japan;
	email: nagahara@ieee.org. D.\ Quevedo is with
	School of Electrical Engineering \& Computer Science,
	The University of Newcastle, Callaghan, NSW 2308,
	Australia; email: dquevedo@ieee.org. J.\ {\O}stergaard is with
        Department of Electronic Systems, 
	Multimedia Information and Signal Processing (MISP),
	Aalborg University,
	Niels Jernes Vej 12, A6-305,
	DK-9220, Aalborg; email: jo@es.aau.dk.}%
	
\thanks{This research was supported in part under 
	the JSPS Grant-in-Aid for Scientific Research (C) No.~24560543,
	and also
	Australian Research Council's
	Discovery Projects funding scheme (project number DP0988601).} 
}

\maketitle

\begin{abstract}
We study feedback control
over erasure channels with packet-dropouts.
To achieve robustness with
respect to packet-dropouts, the controller transmits  data
packets  
containing plant input predictions, which minimize a finite horizon
cost function.
To reduce the data size of packets,
we propose to adopt sparsity-promoting optimizations,
namely, $\elll$ and  $\ell^2$-constrained $\ell^0$ optimizations, 
for which efficient algorithms exist.
We show how to design the tuning parameters to ensure (practical) 
stability of the resulting feedback control systems when the number of consecutive 
packet-dropouts is bounded.
\end{abstract}

\section{Introduction}
\label{sec:introduction}
In networked control systems (NCSs) communication between controller(s) and
plant(s) is made through
unreliable and rate-limited communication links
such as wireless networks and the Internet; see e.g.,
\cite{naifag07}.
Many interesting
challenges arise and successful NCS design
methods need to 
consider both control and communication aspects. In particular,
so-called {\em packetized predictive control} (PPC) has been shown to have
favorable stability and performance properties, especially in the presence of packet-dropouts
\cite{Bem98,QueNes11,QueOstNes11}.
In PPC,
the controller output is obtained through minimizing 
a finite-horizon cost function on-line and in a  receding horizon manner.
Each control {\em packet}  contains a sequence
of tentative plant inputs for a finite horizon of
future time instants and is transmitted through a communication channel. 
Packets which are successfully received at
the plant actuator side, are stored in a buffer to be 
used whenever later packets are dropped. 
When there are no
packet-dropouts, PPC reduces to model predictive control.
For PPC to give desirable 
closed-loop properties,  the more unreliable the network is, the larger 
the horizon length (and thus the number of tentative plant input values
contained in each packet) needs to be chosen. Clearly, in principle, this would require increasing
the network bandwidth (i.e., its bit-rate), unless the transmitted signals are suitably
encoded. 
It is well-known that there exists a minimum bit-rate for achieving  stability
of a networked feedback control system
\cite{TatMit04,naifag07}.
The optimal quantizer for the minimum bit-rate is a dynamic vector quantizer,
and is, thus,  hard to use in
many applications.
As an alternative, memoryless scalar quantizers, will often be preferable.
In this case, {\em sparse representations} \cite{Ela} can be used
to reduce the data size of transmitted vectors in PPC.
Sparse representations aim at designing sparse vectors,
which have few non-zero coefficients,
along with optimizing some performance indices.
Since sparse vectors
contain many zero-valued elements, they can be easily
compressed by only encoding a few nonzero coefficients and their locations
with a memoryless scalar quantizer.
Well-known examples of this kind of encoding are JPEG
in image processing \cite{Wal91}
and algebraic CELP in speech coding
\cite[Section 17.11.1]{BenSonHua}. 
Over the past few years, a number of studies have been published
which deal with sparsity for
control, including topics such as
trajectory generation \cite{OhlGusLjuBoy10},
state observation \cite{WakSanVin10,BhaBas11},
optimal control 
\cite{KosKosFed10,SchEbeAll11,FarFuJov11,MatPasMai12,GalMac12,GooHaiQueWel04,QueSilNes08},
and also
sampled-data control \cite{Nag+11,NagQueMatHay12,NagMatHay12}.

The purpose of the present work is to introduce sparsity-promoting optimizations for
networked control with dropouts. We will show that sparsity-promoting cost
functions can be used in PPC to achieve good control performance (as measured by a weighted quadratic
norm of the system state), whilst   transmitting sequences with only few non-zero elements.  By studying the sequence of optimal cost
functions \emph{at the instances of successful reception}, we
derive sufficient conditions for (practical) closed-loop stability
in the presence of bounded packet-dropouts.

\par It is well-known that sparsity-promoting optimization,
which is often described in terms of the so-called $\ell^0$ norm \cite{Ela},
is in principle hard to solve due to its combinatorial nature
\cite{Nat95}.
However, there exist efficient methods 
that compute the solution (or an approximation)
of the optimization 
in the field of {\em compressed sensing} (see e.g., \cite{HayNagTan13}).
We will focus on two such methods:
One is
convex relaxation where the $\ell^1$ norm is used in place of
the highly nonconvex $\ell^0$ norm.
This leads to
{\em $\elll$ optimization}
($\ell^1$-regularized $\ell^2$ optimization),
which can be effectively solved with a fast algorithm
called {\em Fast Iterative Shrinkage-Thresholding Algorithm (FISTA)}
\cite{BecTeb09}.
The other approach to obtain sparse solutions is through adoption of
{\em greedy algorithms}.
A greedy algorithm iteratively builds up 
the approximate solution of the $\ell^0$-norm optimization
by updating the support set one by one.
In particular,
{\em Orthogonal Matching Pursuit} (OMP)
\cite{PatRezKri93}
is quite simple and
known to be dramatically faster than exhaustive search. 

\par 
Our present note complements
our recent conference
contribution\cite{NagQue11}, which adopted an $\elll$
optimization for PPC. A limitation of the approach in\cite{NagQue11} is that for
open-loop unstable systems,  asymptotic stability cannot be obtained  in the
presence of bounded  packet-dropouts; the best one can
hope for is   practical stability. Our current paper also complements the
articles \cite{NagQueOst12a,NagQueOst12b}, 
by presenting a  detailed
technical analysis of the scheme, including proofs of results.  

\par 
The remainder of this note is
organized as follows:
Section~\ref{sec:PPC} revises basic elements of packetized
predictive control. 
In Section \ref{sec:design}, 
we show the motivation of sparsity-promoting optimization for PPC,
and formulate the design of the sparse control packets.
In Section \ref{sec:stability}, we study stability of the resultant networked
control system.
A numerical example is included in 
Section \ref{sec:simulation}. Section \ref{sec:conclusion} draws conclusions.

\subsubsection*{Notation}
We write $\N_0$ for $\{0, 1, 2, 3, \ldots\}$. The identity
matrix (of appropriate dimensions) is denoted via $I$. 
For a matrix (or a vector) $A$, $A^\T$ denotes the transpose.
For a vector $\vv=[v_1,\ldots,v_n]^\T\in\R^n$ and a positive definite matrix $P>0$, 
we define
$\|\vv\|_1\eq |v_1|+\dots+|v_n|$,
$\|\vv\|_2\eq \sqrt{\vv^\T \vv}$,
$\|\vv\|_P \eq \sqrt{\vv^\T P \vv}$,
and
$\|\vv\|_\infty\eq \max\{|v_1|,\dots,|v_n|\}$.
The support set $\supp(\vv)$ of vector $\vv$ is defined as
$\supp(\vv)=\{i: v_i\neq 0\}$, and the ``$\ell^0$ norm''
of $\vv$ is defined as $\|\vv\|_0=|\supp(\vv)|$
where $|\cdot|$ denotes the cardinality of a set. Thus, $\|\vv\|_0$ is the number of  nonzero elements in $\vv$.
For any Hermitian matrix $P$, $\lambda_{\max}(P)$ and $\lambda_{\min}(P)$
 denote the maximum and the minimum eigenvalues of $P$, respectively; $\sigma_{\max}^2(P)\eq\lambda_{\max}(P^\T P)$.

\section{Packetized Predictive Networked Control}
\label{sec:PPC}
Let us consider an
unconstrained discrete-time linear time-invariant
 plant model with a scalar input:
\begin{equation}
  \label{eq:plant}
   \vx(k+1)=A\vx(k)+\Bs u(k), \quad k\in\N_0, \quad
   \vx(0)=\vx_0\in\R^n,
\end{equation}
where $\vx(k)\in\R^n$, $u(k)\in\R$.
Throughout this work, we assume that the pair $(A,\Bs)$ is reachable.

\par
\begin{figure}[tbp]
\centering
\includegraphics[width=0.85\linewidth]{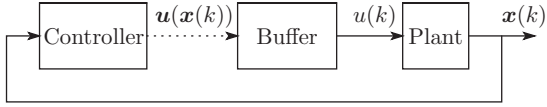}
\caption{Networked Control System with PPC. The dotted line indicates an erasure
  channel.}
\label{fig:NCS}
\end{figure}

 We are interested in an NCS architecture, where the controller
communicates with the plant actuator through an erasure channel, as depicted in
Fig.~\ref{fig:NCS}. 
This channel introduces 
packet-dropouts, which we model via the
dropout sequence $\{d(k)\}_{k \in \N_0}$
where $d(k)=1$ if packet-dropout occurs
and $d(k)=0$ otherwise.
With Packetized Predictive Control (PPC), as
described, for instance, in \cite{QueNes11},
at each time instant $k$, the controller uses the state $\vx(k)$ of the plant
\eqref{eq:plant} to calculate and
sends a control packet of the form
$\vu(\vx(k))=[u_0(\vx(k)),\dots,u_{N-1}(\vx(k))]^\top\in\R^N,$
to the plant input node.

\par To achieve robustness
against packet-dropouts, buffering is used.
More precisely, suppose that at time instant $k$, we have $d(k)=0$, i.e.,
the data packet $\vu(\vx(k))$  is successfully received at
the plant input side. Then, this packet is stored in a buffer, overwriting its
previous contents.
If the next packet $\vu(\vx(k+1))$ is dropped,
then the plant input $u(k+1)$ is set to $u_1(\vx(k))$, the second element of
$\vu(\vx(k))$.
The elements of $\vu(\vx(k))$ are then successively used until
some packet $\vu(\vx(k+\ell))$, $\ell \geq 2$
is successfully received. 

\begin{rem}
It is worth noting that~(\ref{eq:plant}) does not include disturbances. Hence,
as an alternative to PPC,  one could simply transmit the system state to the
actuator and, upon successful reception, the actuator could calculate and
implement a semi-infinite plant input
sequence. In the present work, we focus on situations where the actuator does
not have sufficient computational capabilities precluding such an open-loop control
scheme. In contrast, the sparse PPC formulations proposed in the present work provide feedback at all  instances where no dropouts
occur. Our recent results concerning  related schemes, see \cite{QueNes11,QueNes12}, suggest
that, in the presence of disturbances, PPC will exhibit favorable robustness properties.
\end{rem}

\section{Design of Sparse Control Packets}
\label{sec:design}
In the present section we present two methods for the design of sparse PPC. The purpose is to obtain many zero elements $u_i(\vx(k))$ in the control packet $\vu(\vx(k))$, cf., \cite{GooHaiQueWel04,QueSilNes08}.
The control packet $\vu(\vx(k))$ is designed at each time $k$
via a standard model predictive control formulation:
\begin{equation}
  \vu(\vx(k)) = \argmin_{\vu\in\F(\vx(k))} \bigl\{F(\vx'_N) + \sum_{i=0}^{N-1} L(\vx'_i,u_i)\bigr\},
 \label{eq:general_optimization}
\end{equation} 
where $\vx'_i$ and $u_i$ are state and input predictions, respectively, defined by
$\vx'_0 = \vx$, $\vx'_{i+1} = A\vx'_i + \Bs u_i$,
and $\vu=[u_0,u_1,\dots,u_{N-1}]^\top$.
The function $F$ defines the \emph{terminal cost} and $L$, the \emph{stage cost}.
In~\eqref{eq:general_optimization}, we have introduced a constraint set
$\F(\vx)$, which is assumed to be a closed subset of $\R^N$ and allowed to depend on the state
observation $\vx$. More details on the choice of $\F(\vx)$ will be given below.

To obtain a sparse control vector $\vu(\vx(k))$, we will investigate 
two types of sparsity-promoting optimizations, namely, unconstrained $\elll$
optimization and
$\ell^2$-constrained $\ell^0$ optimization.

\subsection{Unconstrained $\elll$ optimization}
Here, the terminal and stage costs are given by
$L(\vx,u) = \|\vx\|_Q^2 + \mu \left|u\right|$
and  $F(\vx) = \|\vx\|_P^2$,
where $\mu>0$, $P>0$, and $Q>0$,
and  $\F(\vx)$ is taken as $\R^N$ (unconstrained).
With the following notation:
\begin{equation}
 \begin{split}
  G &\eq \bar{Q}^{1/2}\Phi,~ H \eq -\bar{Q}^{1/2}\Upsilon,~
  \bar{Q} \eq \mathrm{diag}\{Q,\ldots,Q, P\},\\
  \Phi &\eq \begin{bmatrix}\Bs & 0 & \ldots & 0\\
  			A\Bs & \Bs & \ldots & 0\\
			\vdots & \vdots & \ddots & \vdots\\
			A^{N-1}\Bs & A^{N-2}\Bs & \ldots & \Bs
			\end{bmatrix},\quad
  \Upsilon \eq \begin{bmatrix}A\\A^2\\\vdots\\A^N\end{bmatrix},
 \end{split}
 \label{eq:matrix-defn}
\end{equation}
the optimization can be represented by
\begin{equation}
\label{eq:opt1}
\begin{split}
  \vu(\vx) &= \argmin_{\vu\in\R^N}J(\vx,\vu)\\
J(\vx,\vu)&\eq\|G\vu-H\vx\|_2^2+\mu\|\vu\|_1+\|\vx\|_Q^2.
\end{split}
\end{equation}
This   {\em (unconstrained) $\elll$ optimization} is known to produce a sparse vector
via very efficient algorithms 
\cite{HayNagTan13}.
By simulation, we often obtain much sparser vectors than those produced by
the $\ell^0$ method described below (see also Section \ref{sec:simulation}).
However, due to fundamental properties of the $\elll$ optimization,
the state $\vx(k)$ never converges to the origin if the plant \eqref{eq:plant}
is unstable  (see Proposition \ref{prop:Omega} below).
Instead, in Theorem \ref{thm:stability_l1l2} we will establish that, if design
parameters $P>0$ and $\mu>0$ 
in \eqref{eq:matrix-defn} and \eqref{eq:opt1} are appropriately chosen,
then the state
$\vx(k)$ converges into a closed finite set including the origin.

\subsection{$\ell^2$-Constrained $\ell^0$ optimization}
Here,  $F(\vx)=0$, whereas the
stage cost $L$ is given by
$L(\vx,u) = 1$ if $u\neq0$ and $L(\vx,u) = 0$ if $u=0$.
The constraint set $\F(\vx)$, used in~\eqref{eq:general_optimization} is taken as
\begin{equation}
 \F(\vx) \eq \bigl\{\vu\in\R^N: \|\vx'_N\|_P^2 + \sum_{i=1}^{N-1}\|\vx'_i\|_Q^2 \leq \|\vx\|_W^2\bigr\},
 \label{eq:feasible_set_l0}
\end{equation}
where $\vx'_1,\dots,\vx'_N$ are predicted states that depend on $\vu$.
We assume $P>0$, $Q>0$, and $W>0$ are chosen such that the set $\F(\vx)$ is non-empty for all $\vx\in\R^N$.
In terms of the  notation \eqref{eq:matrix-defn},
the optimization can, thus, be written as
\begin{equation}
 \begin{split}
  \vu(\vx) &= \argmin_{\vu\in\F(\vx)} \|\vu\|_0,\\
  \F(\vx) &= \{\vu\in\R^N: \|G\vu-H\vx\|_2^2\leq \|\vx\|_W^2\}.
 \end{split}
 \label{eq:opt0}
\end{equation}

The $\ell^0$ optimization \eqref{eq:opt0}
is in general extremely difficult to solve
since it requires a combinatorial search
that explores all possible sparse supports of $\vu\in\R^N$.
In fact, it has been proven to be NP hard \cite{Nat95}, leading to the
development of sub-optimal algorithms; see, e.g.
\cite{HayNagTan13}.
One approach to the combinatorial optimization is
an iterative greedy algorithm
called {\em Orthogonal Matching Pursuit} (OMP)
\cite{PatRezKri93}.

\begin{rem}
\label{rem:sparsity-constrained}
In \cite{GooHaiQueWel04,ImeBas06,She12,ShiYuaChe12},  sparse control methods for
closed loop control without dropouts were studied, in essence, adopting an {\em $\ell^0$-constrained (or sparsity-constrained)
$\ell^2$ optimization}:
\[
  \vc{u}(\vc{x}) = \argmin_{\vc{u}\in\F} \|G\vc{u}-H\vc{x}\|_2^2,~
  \F = \{\vc{u}\in\R^N: \|\vc{u}\|_0\leq S\},
\]
where $S$ is a positive integer less than $N$. The optimization above may be effectively solved via the CoSaMP algorithm
described in \cite{NeeTro08}.
Since the bound $S$ of $\|\vc{u}\|_0$  is specified {\em a priori}, one can adopt the interleaved single pulse permutation
(ISPP) design
\cite[Section 17.11.1]{BenSonHua} for effectively encoding the support data of $\vc{u}$.
A disadvantage of this approach is the difficulty in estimating a bound 
$S$ that guarantees stability of the feedback loop. 
In contrast, in the following section we will show how design parameters 
in \eqref{eq:general_optimization}
can be chosen to ensure closed loop stability in the presence of bounded dropouts.
\end{rem}

\section{Stability Analysis of Sparse PPC Loops}
\label{sec:stability}
In this section, we provide stability results for
sparse PPC with $\elll$ optimization and $\ell^2$-constrained $\ell^0$
optimization, as presented in Section \ref{sec:design}.
To establish deterministic stability properties,
we here impose a bound on the maximum number of successive dropouts as
follows:\footnote{Such a deterministic bound may arise for example in
wired control networks with $N$ users, if the sole cause of dropouts is contention between users.
When a priority-based,
deterministic re-transmission protocol is used, transmission is then guaranteed within $N$
intervals. Alternatively, in a wireless scenario with random packet-dropouts, more than $N$
consecutive packet drops might trigger a hypothesis that there is a fault in the network,
requiring a higher-level network response outside the present formulation.}
\begin{ass}[Bounded packet-dropouts]
\label{ass:dropouts}
The number of
consecutive packet-dropouts is uniformly bounded by $N-1$.
\hfs
\end{ass}
In view of the above, the horizon length $N$ in~(\ref{eq:general_optimization}) allows one to trade  computational
  complexity of the on-line optimization for robustness with respect to
  dropouts. Thus, the less  reliable the network is, the larger $N$
  should be chosen.

\subsection{Stability Analysis of $\elll$ PPC}
\label{subsec:stability_l1l2}

We here analyze closed-loop stability of  $\elll$ PPC, see  \eqref{eq:opt1}.
Our
analysis uses elements 
of the technique introduced in 
\cite{QueNes11}.
A distinguishing aspect of the situation at hand is that, for open-loop unstable
plants, even when there are no packet-dropouts, asymptotic stability will not be
achieved,  despite the fact that the plant-model in \eqref{eq:plant} is
disturbance-free. 
In fact, we have the following proposition,
which is directly proved from an equivalent dual problem of
\eqref{eq:opt1} (see \cite[Appendix B]{Fuc01}).
\begin{prop}
\label{prop:Omega}
Define
$\Omega\eq \{\vx\in\R^n: \|G^\T H\vx\|_\infty\leq \mu/2\}.$
If $\vx\in\Omega$, then $\vu(\vx)=\vz$.\hfill $\square$
\end{prop}

It follows that if $\vx(k)\in \Omega$ and there are no dropouts at time $k$, then
the control will be $u(k)=0$.
That is, the control system \eqref{eq:plant} behaves as an open-loop system in the set $\Omega$.
Hence, asymptotic stability will in general not be achieved, if $A$
has eigenvalues outside the unit circle.
This fundamental property is linked to
sparsity of the control vector.

By the fact mentioned above, we will next turn our attention to {\em practical} stability (i.e.,
stability of a set) of the associated
networked control system. For that purpose, we will analyze the value
function 
\begin{equation}
  V(\vx) \eq \min_{\vu\in\R^N} J(\vx,\vu),
   \label{eq:Vx}
\end{equation}
where $J(\vx,\vu)$ is as in~(\ref{eq:opt1}).
 First, we find   bounds of $V(\vx)$.
\begin{lem}[Bounds of $V$]
\label{lem:V_bounds}
For any $\vx\in\R^n$, we have
\[
  \lambda_{\min}(Q)\|\vx\|_2^2 \leq V(\vx) \leq \phi(\|\vx\|_2),
\]
where
$\phi(t)\eq a_1t + (a_2+\lambda_{\max}(Q))t^2$,
$a_1 \eq \mu \sqrt{n}~ \sigma_{\max}\left(G^\dagger H\right)$,
$a_2 \eq \lambda_{\max} (W^\star)$,
and the matrices $G^\dagger$ and $W^\star$ are
given by
\begin{equation}
 G^\dagger\eq (G^\top G)^{-1}G^\top,\quad
 W^\star \eq H^\top (I-GG^\dagger)H.
 \label{eq:Wstar}
\end{equation}

\end{lem}
\begin{IEEEproof}
Applying $\vu^\star(\vx)=G^\dagger H\vx$ to the cost in \eqref{eq:Vx} gives
	\[
	 \begin{split}
	    &\|(GG^\dagger-I)H\vx\|_2^2 + \mu \|G^\dagger H \vx\|_1 +  \|\vx\|_Q^2\\
	   &~\leq \mu \sqrt{n} \sigma_{\max}(G^\dagger H)\|\vx\|_2
	   + (\lambda_{\max}(W^\star)+\lambda_{\max}(Q))\|\vx\|_2^2,
	 \end{split}
	\] 
where we used the norm inequality $\|\vv\|_1 \leq \sqrt{n}\|\vv\|_2$
for any $\vv\in\R^n$, see \cite{Ber}.
This and the definition of $V(\vx)$ in \eqref{eq:Vx} provide the upper bound on $V(\vx)$.
To obtain the lower bound, we simply note that by the definition of $J(\vx,\vu)$, we have
$\|\vx\|_Q^2 \leq J(\vx,\vu)$
for any $\vu\in\R^N$, and hence $\lambda_{\min}(Q)\|\vx\|_2^2\leq V(\vx)$.
\end{IEEEproof}

\begin{rem}
\label{rem:LS_approximation}
The least squares solution $\vu^\star(\vx) = G^\dagger H\vx$ 
approximates the optimizer $\vu(\vx)$ of the $\elll$
cost in \eqref{eq:opt1} for given plant state $\vx$. 
Since the following bound holds \cite{Fuc04}:
\[
  \|G^\T(G\vu(\vx) - H\vx)\|_\infty = \|G^\T G(\vu(\vx) - \vu^\star(\vx))\|_\infty\leq\mu/2,
\]
the upper bound for $V(\vx)$ given in Lemma~\ref{lem:V_bounds} will be tight, if
$\mu$ is small.\hfs
\end{rem}

Having established the above preliminary results, 
we introduce the  $i$-th iterated mapping $\vf^{i}$
with the optimal vector $\vu(\vx)=[u_0(\vx),\dots,u_{N-1}(\vx)]^\top$
defined in \eqref{eq:opt1} through the recursion
\[
 \vf^{i}(\vx) \eq A^i\vx + \sum_{l=0}^{i-1} A^{i-1-l}Bu_l(\vx),\quad i=1,2,\ldots, N.
\]
This mapping describes the plant state evolution during periods of consecutive packet-dropouts. 
Note that, since the input $\vu(\vx)$ is not a linear function of $\vx$ (see Proposition \ref{prop:Omega}), 
the function $\vf^{i}(\vx)$ is nonlinear.
The following bound plays a crucial
role to establish deterministic stability guarantees:
 \begin{lem}[Open-loop bound]
\label{lem:openloop_bound}
Assume that $P>0$ satisfies the following Riccati equation
\begin{equation}
 P = A^\top P A - A^\top P\Bs(\Bs^\top P\Bs+r)^{-1}\Bs^\top PA + Q
 \label{eq:ric}
\end{equation}
with
$r = \mu^2N/(4\epsilon)$, $\epsilon>0$.
Then for any $\vx\in\R^n$ and $i=1,2,\dots,N$, we have
\begin{equation}
 V(\vf^{i}(\vx))-V(\vx) \leq -\lambda_{\min}(Q) \|\vx\|_2^2 + \epsilon.
 \label{eq:open-loop-bound}
\end{equation}
\end{lem}
\begin{IEEEproof}
Fix $i\in\{1,\ldots,N-1\}$ and consider the sequence
\[
 \tilde{\vu} = \left\{u_i(\vx), u_{i+1}(\vx), \dots, u_{N-1}(\vx), \tilde{u}_N, \ldots, \tilde{u}_{N+i-1}\right\},
\]
where $\tilde{u}_{N+j}$ ($j=0,1,\ldots,i-1$) is given by
$\tilde{u}_{N+j} = K\tilde{\vx}_{N+j}$ and
$\tilde{\vx}_{N+j+1} = A\tilde{\vx}_{N+j} + B \tilde{u}_{N+j}$,
where
$K = -(B^\top PB+r)^{-1}B^\top PA$
and $\tilde{\vx}_N = \vf^N(\vx)$.
We then have
\[
 \begin{split}
   J(\vf^{i}(\vx),\tilde{\vu})
    &= V(\vx) - \sum_{l=0}^{i-1}\left\{\|\vf^l(\vx)\|_Q^2 + \mu \left| u_l(\vx)\right| \right\}\\
	&~+ \sum_{l=N}^{N+i-1} \left\{\|\tilde{\vx}_{l+1}\|_P^2 - \|\tilde{\vx}_l\|_P^2 + \|\tilde{\vx}_l\|_Q^2 + \mu\left| \tilde{u}_l(\vx)\right| \right\}.
 \end{split}
\]
By the relation $\tilde{\vx}_{l+1} = (A+BK)\tilde{\vx}_l$ and $\tilde{u}_l = K\tilde{\vx}_l$ for $l=N,N+1,\ldots, N+i-1$,
we can bound the terms in the last sum above by
\[
 \begin{split}
 &\|\tilde{\vx}_{l+1}\|_P^2 - \|\tilde{\vx}_l\|_P^2 + \|\tilde{\vx}_l\|_Q^2 + \mu \left| \tilde{u}_l(\vx)\right|\\
  &=\tilde{\vx}_l^\T\!\left[(A+BK)^\T P(A+BK) - P + Q + \frac{\mu^2 N}{4\epsilon}K^\T K\right]\!\tilde{\vx}_l\\
	&\quad - \frac{\mu^2N}{4\epsilon}\left(|K\tilde{\vx}_l|-\frac{2\epsilon}{\mu N}\right)^2 + \frac{\epsilon}{N} \leq \frac{\epsilon}{N}.
 \end{split}
\]
Thus, the cost function $J(\vf^{i}(\vx),\tilde{\vu})$ can be upper
bounded by
\begin{equation}
 \begin{split}
	  J(\vf^{i}(\vx),\tilde{\vu})
	  &\leq V(\vx) - \lambda_{\min}(Q) \|\vx\|_2^2 + \epsilon,
 \end{split}
 \label{eq:J_bound}
\end{equation}
where we have used the relation $\vf^0(\vx) = \vx$.
Since $V(\vf^i(\vx))$ is the minimal value of $J(\vf^i(\vx),\vu)$ among all $\vu$'s in $\R^N$,
we have $V(\vf^i(\vx))\leq J(\vf^i(\vx),\tilde{\vu})$ , and hence
inequality \eqref{eq:open-loop-bound} holds.
For the case $i=N$, we consider the sequence
$\tilde{\vu} = \left\{\tilde{u}_N, \tilde{u}_{N+1}, \ldots, \tilde{u}_{2N-1}\right\}$.
If we define $\sum_{l=N}^{N-1} = 0$,
then \eqref{eq:J_bound} follows as in the case $i\leq N-1$.
\end{IEEEproof}

The above result can be used to derive the following contraction property of the
optimal costs during periods of successive packet-dropouts:
\begin{lem}[Contractions]
\label{lem:contracting}
Let $\epsilon>0$. Assume that $P>0$ satisfies \eqref{eq:ric} with
$r = \mu^2N/(4\epsilon)$.
Then there exists a real number $\rho\in (0,1)$ such that
for all $\vx\in\R^n$, we have
\[
 V(\vf^i(\vx)) \leq \rho V(\vx) + \epsilon + \lambda_{\min}(Q)/4,\quad i=1,2,\ldots, N.
\] 
\end{lem}
\begin{IEEEproof}
In this proof, we borrow a technique used in the proof of \cite[Theorem 4.2.5]{Laz}.
By Lemma \ref{lem:V_bounds}, for $\vx\neq \vz$ we have
$$0 < V(\vx) \leq a_1 \|\vx\|_2 + (a_2 + \lambda_{\max}(Q)) \|\vx\|_2^2.$$
 Now suppose that $0<\|\vx\|_2\leq 1$.
Then $\|\vx\|_2^2 \leq \|\vx\|_2$ and hence
$V(\vx) \leq (a_1 + a_2 + \lambda_{\max}(Q)) \|\vx\|_2$.
From Lemma \ref{lem:openloop_bound}, it follows that
\[
 \begin{split}
  V(\vf^i(\vx))
  &\leq (1-\lambda_{\min}(Q)\|\vx\|_2V(\vx)^{-1})V(\vx)\\
	&\qquad-\!\lambda_{\min}(Q)\!\left(\|\vx\|^2_2-\|\vx\|_2\right) + \epsilon\\
  &\leq \rho V(\vx) + \lambda_{\min}(Q)/4 + \epsilon,
 \end{split}
\]
with
\begin{equation}
 \rho \eq 1-\lambda_{\min}(Q)(a_1+a_2+\lambda_{\max}(Q))^{-1}.
 \label{eq:rho_l1l2}
\end{equation}
Since $0<\lambda_{\min}(Q)\leq\lambda_{\max}(Q)$, $a_1>0$, and $a_2>0$,
it follows that $\rho \in (0,1)$.

\par Next,  consider the case where $\|\vx\|_2>1$ so that
 $\|\vx\|_2 < \|\vx\|_2^2$ and
$V(\vx) < (a_1 + a_2 + \lambda_{\max}(Q))\|\vx\|_2^2$.
This and  Lemma \ref{lem:openloop_bound} give
\[
 \begin{split}
  V(\vf^i(\vx))
   &<\rho V(\vx) + \epsilon
   \leq \rho V(\vx) + \lambda_{\min}(Q)/4 + \epsilon.
 \end{split}  
\]
If $\vx=\vz$, then the above inequality also holds since $V(\vz)=\vz$.
\end{IEEEproof}

We will next use Lemma~\ref{lem:contracting} to establish sufficient conditions
for 
practical
stability of $\elll$ PPC
in the presence of packet-dropouts.  
Theorem~\ref{thm:stability_l1l2} stated below shows how to design the parameters of
the cost function to ensure practical stability  
in the presence of bounded packet-dropouts satisfying Assumption~\ref{ass:dropouts}.
\begin{thm}[Practical stability of $\ell^1$-$\ell^2$ PPC]
\label{thm:stability_l1l2}
Let $\epsilon>0$ and choose $P>0$ to satisfy \eqref{eq:ric} with 
$r = \mu^2N/(4\epsilon)$.
Then $\|\vx(k)\|_2$ is bounded for all $k\in\N_0$, and
\begin{equation}
  \lim_{k\rightarrow\infty} \|\vx(k)\|_2 \leq R\eq \sqrt{\frac{1}{1-\rho}\left(\frac{\epsilon}{\lambda_{\min}(Q)}+\frac{1}{4}\right)},
  \label{eq:thm_l1l2}
\end{equation}
where $\rho$ is given in
\eqref{eq:rho_l1l2}.

\end{thm}

\begin{IEEEproof}
Denote the
time instants where there are no
packet-dropouts, i.e., where
$d(k)=0$, as
\begin{equation}
  \mathcal{K}\eq\{k_i\}_{i\in\N_0}\subseteq \N_0, \quad
  k_{i+1}>k_i,\; \forall i \in \N_0,
  \label{eq:nodropouts}
\end{equation}
whereas the number of consecutive packet-dropouts is denoted via:
\begin{equation}
  m_i\eq k_{i+1}-k_i-1,\quad i\in\N_0.
  \label{eq:mi}
\end{equation}
Note that $m_i \geq 0$,
with equality if and only if no dropouts occur between instants $k_i$ and
$k_{i+1}$.
Fix $i\in\N_0$ and note that at time instant $k_i$, the control packet is successfully transmitted to the buffer.
Then until the next packet is received at time $k_{i+1}$, $m_i$ consecutive packet-dropouts occur.
By the PPC strategy,  the control input becomes 
$u(k_i+l) = u_l(\vx(k_i))$, $l=1,2,\ldots,m_i$,
and, since \eqref{eq:plant} is exact,
the states $\vx(k_i+1),\ldots,\vx(k_i+m_i)$
are determined by these open-loop controls.
Since we have $m_i\leq N-1$ from Assumption \ref{ass:dropouts},
Lemma \ref{lem:contracting} gives
\begin{equation}
  V(\vx(k)) 
   \leq \rho V(\vx(k_i)) + \epsilon + \lambda_{\min}(Q)/4,
   \label{eq:thm_l1l2_1}
\end{equation} 
for $k\in\{k_i+1, k_i+2,\ldots, k_i+m_i\}$, and also for $k_{i+1}=k_i+m_i+1$, we have
\begin{equation}
 V(\vx(k_{i+1})) \leq \rho V(\vx(k_i)) + \epsilon + \lambda_{\min}(Q)/4.
 \label{eq:thm_l1l2_2}
\end{equation}
Now by induction from \eqref{eq:thm_l1l2_2}, it is easy to see that
from Lemma \ref{lem:V_bounds},
\[
	 \begin{split}
	  V(\vx(k_i))
	   &\leq \rho^i V(\vx(k_0))
		+ (1+\cdots+\rho^{i-1})(\epsilon + \lambda_{\min}(Q)/4)\\
	   &\leq \rho^i \phi(\|\vx(k_0)\|_2) + (1-\rho)^{-1}(\epsilon + \lambda_{\min}(Q)/4).
	 \end{split}
\]
This inequality and \eqref{eq:thm_l1l2_1} give the bound
\[
 \begin{split}
  V(\vx(k)) 
   &\leq \rho^{i+1} \phi(\|\vx(k_0)\|_2) + (1-\rho)^{-1} (\epsilon + \lambda_{\min}(Q)/4),\\
 \end{split}
\]
for $k\in\{k_i+1, k_i+2,\ldots, k_{i+1}-1\}$, and this inequality also holds for $k=k_{i+1}$.
Finally, by using the lower bound of $V(\vx)$ provided in Lemma \ref{lem:V_bounds}, we have
\[
 \begin{split}
  \|\vx(k)\|_2
   &\leq \sqrt{\frac{V(\vx(k))}{\lambda_{\min}(Q)}}
   \leq (\sqrt{\rho})^{i+1} \sqrt{\frac{\phi(\|\vx(k_0)\|_2)}{\lambda_{\min}(Q)}}+R,\\
 \end{split}
\]
where $R$ is defined in \eqref{eq:thm_l1l2} and we used 
the inequality $\sqrt{a+b} \leq \sqrt{a} +\sqrt{b}$, for all $a,b\geq 0$.
The above inequality leads to   \eqref{eq:thm_l1l2}.
\end{IEEEproof}

Theorem \ref{thm:stability_l1l2} establishes practical stability of the networked
control 
system. It shows that, provided the conditions are met, the plant state will be
ultimately bounded in a  ball of radius $R$.
It is worth noting that, as in other stability results which use Lyapunov
techniques, this bound will, in general, not be tight.

\subsection{Stability Analysis of $\ell^0$ PPC}
\label{subsec:stability_l0}

Here we analyze closed-loop stability of
$\ell^0$ PPC, as described in \eqref{eq:opt0}. Since stability will unavoidably be
linked to feasibility, we begin with analysis 
of the feasible set $\F(\vx)$ given in \eqref{eq:feasible_set_l0}.
Clearly, for given matrices $G$ and $H$, the feasible set $\F(\vx)$ will be non-empty
if the matrix $W$ is ``larger'' than $W^\star$ given in \eqref{eq:Wstar},
the ``smallest'' $W$ which ensures that  $\F(\vx)\not = \emptyset$.
In fact, we have: 
\begin{lem}
\label{lem:W}
Let
$$\F^\star(\vx) \eq \{\vu\in\R^N: \|G\vu-H\vx\|_2^2\leq \|\vx\|_{W^\star}^2\}.$$
For any $W\geq W^\star$,
we have $\F(\vx)\supseteq\F^\star(\vx)$.
Moreover, if $W\geq W^\star$,
then the feasible set $\F(\vx)$
is a closed, convex, and non-empty subset of $\R^N$.
\end{lem}
\begin{IEEEproof}
Suppose $W\geq W^\star$. Then $\vu^\star(\vx)\in\F(\vx)$ and hence $\F(\vx)\supseteq\F^\star(\vx)$.
This also implies that $\F(\vx)$ is non-empty for any $W\geq W^\star$.
Closedness and convexity of $\F(\vx)$ are obvious since it is defined by a quadratic form
(the set is a closed ellipsoid in $\R^N$).
\end{IEEEproof}

Based on this lemma, we hereafter assume that
\begin{equation}
 \E \eq W-W^\star>0.
 \label{eq:Eps}
\end{equation}
The feasible solutions for~\eqref{eq:opt0} can be characterized as follows:
\begin{lem}[Feasible Solutions]
\label{lem:eps_bound}
For any $\vu\in\F(\vx)$, there exists $\eps(\vx)\in\R^N$ such that
\begin{equation}
\label{eq:Ueps}
 \vu=\vu^\star(\vx) + \eps(\vx),~\text{with}\quad\|G \eps(\vx)\|^2_2 \leq \|\vx\|_\E^2,
\end{equation}
where $\E$ is given in \eqref{eq:Eps}.
\end{lem}
\begin{IEEEproof}
The fact $G\eps(\vx)\perp(G\vu^\star(\vx)-H\vx)$ gives the result.
\end{IEEEproof}
\begin{rem}
\label{rem:penalty}
The error term $\eps(\vx)$ in~\eqref{eq:Ueps} may be interpreted as a
``penalty charge'' for sparsifying the vector (control packet) $\vu$,
since the term $\|G\vu(\vx)-H\vx\|_2^2$ with the sparse control $\vu(\vx)$
will be larger than with the least squares one,
$\|G\vu^\star(\vx)-H\vx\|_2^2$. \hfill$\square$
\end{rem}

Now take $Q>0$ arbitrarily and let $P>0$ be
the solution to the Riccati equation \eqref{eq:ric}
with $r=0$. Then,
from Lemma \ref{lem:eps_bound}
and well-known results in dynamic programming
\cite[Chapter 3]{Bert},
all feasible control vectors
$\vu(\vx)\in\F(\vx)$
can be written as
\[
 u_i(\vx) = \K(A+\Bs\K)^i\vx + \varepsilon_i(\vx),\quad i=0,1,\ldots, N-1,
\]
where 
\begin{equation}
 \K \eq -(\Bs^\top P\Bs)^{-1}\Bs^\top PA,
 \label{eq:K}
\end{equation}
and $\varepsilon_i(\vx)$ is the $(i+1)$-th element of $\eps(\vx)$
satisfying the inequality in \eqref{eq:Ueps}.
The associated open-loop states are
\begin{equation}
  \vx_0 = \vx,\quad
  \vx_{i+1} = (A+\Bs\K)\vx_i + \Bs w_i(\vx),
 \label{eq:trajectory}
\end{equation}
where
\begin{equation}
 w_i(\vx) \eq \varepsilon_i(\vx) - \K\sum_{l=0}^{i-1} A^{i-1-l}\Bs\varepsilon_l(\vx).
 \label{eq:wi}
\end{equation}
By using the definition \eqref{eq:K} of the matrix $K$,
we have
\[
  w_i(\vx) 
   = (\Bs^\top P\Bs)^{-1}\Bs^\top P\Phi_i\eps(\vx),\quad i=0,1,\ldots, N-1,
\]
where $\Phi_i\in\R^{n\times N}$ is the $(i+1)$-th row block of the matrix $\Phi$
defined in \eqref{eq:matrix-defn}.

\par For the $\ell_0$-case, we use the  quadratic function
$V_P(\vx) \eq \|\vx\|_P^2$
as a Lyapunov function candidate for the system at the times of successful
transmission instants.
The following result bears some similarities to  Lemma~\ref{lem:contracting},
with the important difference that in~\eqref{eq:openloop_stability2}  the upper bound
goes to zero as $\vx $ goes to the origin.  
\begin{lem}[Contractions]
\label{lem:openloop_bound_l0}
Suppose $Q>0$ is chosen arbitrarily, $P>0$ is the solution of 
the Riccati equation \eqref{eq:ric}
with $r=0$,
and $W>0$ is such that $W>W^\star$.
Let $\E=W-W^\star$.
Then there exist constants $\rho\in[0,1)$ and $c>0$ such that
\begin{equation}
V_P(\vx_i) \leq \rho^i V_P(\vx) + c\|\vx\|_{\E}^2,\quad i=1,2,\ldots,N.
\label{eq:openloop_stability2}
\end{equation}
\end{lem}

\begin{IEEEproof}
  Substitution of 
the state $\vx_{i+1}$ given in \eqref{eq:trajectory}
into $V_P(\vx)$
yields that
$$V_P(\vx_{i+1}) = V_P(\vx_i) -\|\vx_i\|_Q^2 + \Bs^\top P\Bs\left|w_i(\vx)\right|^2.$$
By the definition of $w_i(\vx)$ in \eqref{eq:wi}, we have
\[
 \begin{split}
  (\Bs^\top P\Bs)&\left|w_i(\vx)\right|^2
   \leq\eps(\vx)\Phi_i^\top P\Phi_i\eps(\vx)\\
   &\leq \max_{i}\lambda_{\max}\bigl\{\Phi_i^\top P\Phi_i(G^\top G)^{-1}\bigr\}
   \|G \eps(\vx)\|_2^2 \leq c_1\|\vx\|_{\E}^2,
 \end{split}
\]
where the last inequality is due to Lemma \ref{lem:eps_bound}, and 
\begin{equation}
 c_1\eq \max_{i=0,\ldots,N-1} \lambda_{\max} \bigl\{\Phi_i^\top P\Phi_i(G^\top G)^{-1}\bigr\}>0.
 \label{eq:c1}
\end{equation}
Then the above inequality gives
\begin{align}
  V_P(\vx_{i+1})
   &\leq (1-\|\vx_i\|_Q^2\|\vx_i\|_P^{-2})V_P(\vx_i) + c_1\|\vx\|_{\E}^2\nonumber\\
   &\leq \rho V_P(\vx_i) + c_1\|\vx\|_{\E}^2,\quad
   \rho \eq 1-\lambda_{\min}(QP^{-1}). \label{eq:rho_l0}
\end{align}
Since $P\geq Q>0$, we have $\rho\in[0,1)$.
By mathematical induction, we finally obtain
\begin{align}
  V_P(\vx_i)
   &\leq \rho^i V_P(\vx) + (\rho^{i-1}+\cdots+\rho+1)c_1\|\vx\|_{\E}^2\nonumber\\
   &\leq \rho^i V_P(\vx) + c\|\vx\|_{\E}^2,
 \label{eq:c}
\end{align}
where $\quad
   c \eq (1-\rho)^{-1}(1-\rho^N)c_1$.
\end{IEEEproof}

Having established the contraction property~\eqref{eq:openloop_stability2}, the
following result shows that $\ell^0$-PPC can be tuned to give asymptotic
stability in the presence of bounded packet-dropouts:
\begin{thm}[Asymptotic Stability]
\label{thm:stability_l0}
Suppose that  the matrices $P$, $Q$, and $W$
are chosen by the following procedure:
\begin{enumerate}
\item Choose $Q>0$ arbitrarily.
\item Solve the Riccati equation \eqref{eq:ric} with $r=0$ to obtain $P>0$.
\item Compute $\rho\in[0,1)$ and $c>0$ via \eqref{eq:c1}, 
\eqref{eq:rho_l0}, and \eqref{eq:c}.
\item Choose $\E$ such that $0<\E<(1-\rho)P/c$.
\item Compute $W^\star=P-Q$ and set $W=W^\star + \E$.
\end{enumerate}
Then the sparse control packets $\vu(\vx(k))$, $k\in\N_0$,
 optimizing \eqref{eq:opt0}
with the above matrices,
lead to asymptotic stability of the networked control system,
that is, $\vx(k)\rightarrow\vz$ as $k\rightarrow\infty$.
\end{thm}
\begin{IEEEproof}
We use the notation of time instants where there are no packet-dropouts
given in \eqref{eq:nodropouts} and \eqref{eq:mi}.
Fix $i\in\N_0$.
Since we have $m_i\leq N-1$ from Assumption \ref{ass:dropouts},
Lemma \ref{lem:openloop_bound_l0} gives
\begin{equation}
  V_P(\vx(k))
   \leq\vx(k_i)^\top \left(\rho P + c\E\right)\vx(k_i)
   <V_P(\vx(k_i)),
 \label{eq:Vxk}
\end{equation}
for $k=k_i,k_i+1,\ldots,k_i+m_i$.
Also, for $k_{i+1}=k_i+m_i+1$,
the next instant when the control packet is successfully transmitted,
we have
$$V_P(\vx(k_{i+1})) < V_P(\vx(k_{i+1}-1)) < V_P(\vx(k_i)).$$
It follows that at the time instants $k_0,k_1,\ldots$ (no-dropout instants), 
$0\leq V_P(\vx(k_i))$ strictly decreases,
and hence
$\vx(k_i) \rightarrow \vz$
as $i\rightarrow\infty$.
Then, by \eqref{eq:Vxk}, 
for $k=k_i, k_i+1,\ldots,k_i+m_i$ (consecutive dropout instants),
$V_P(\vx(k))$ is bounded by $V_P(\vx(k_i))$. Since the latter
converges to zero,  we conclude that $\vx(k)\rightarrow\vz$ as $k\rightarrow\infty$.
\end{IEEEproof}

In summary, the networked control system affected by bounded packet-dropouts
is asymptotically  stable
with the sparse control packets
obtained by the optimization \eqref{eq:opt0}
if $P$, $Q$, and $W$ are computed as per Theorem~\ref{thm:stability_l0}.

\section{Simulation Study}
\label{sec:simulation}

To assess the effectiveness of
the proposed sparse control methods, we consider a plant model of the form 
\eqref{eq:plant}
with%
\footnote{The elements of these matrices are generated by random sampling from
the normal distribution with mean 0 and variance 1.
Note that the matrix $A$ has 2 unstable eigenvalues
($1.5259$ and $-1.1441$)
and 2 stable eigenvalues
($0.4198$ and $-0.7724$).}
{\footnotesize
	\[
	 A = \left[\begin{array}{rrrr}
	    \!\!0.0685&    \!\!1.1221&   \!\!-0.6615&    \!\!0.3087\!\!\\
	    \!\!0.9512&    \!\!0.3237&   \!\!-0.2253&   \!\!-0.5701\!\!\\
	   \!\!-0.3448&   \!\!-0.4112&   \!\!-0.8299&    \!\!0.5388\!\!\\
	    \!\!0.0359&   \!\!-0.6418&   \!\!-0.1262&    \!\!0.4669\!\!\\
	 \end{array}\right],~
	 B = \left[\begin{array}{r}
	    \!\!2.3459\!\!\\
	    \!\!0.0893\!\!\\
	    \!\!2.2103\!\!\\
	    \!\!0.7440\!\!\\
	 \end{array}\right].
	\]
}
We set $N=10$ and
packet-dropouts are simulated with a model
where the probability distribution
of the number of consecutive dropouts is uniform over $[1,\dots,N-1]$.
For the system above, we make simulation-based examination of
the proposed stabilizing PPC formulations using sparsity-promoting optimizations,
$\elll$ optimization with FISTA and $\ell^0$ optimization
with OMP.
We set the  horizon length (or the packet size) to $N=10$.

\par To compare these two sparsity-promoting methods with traditional PPC approaches,
we also consider
the $\ell^2$-optimal control $$\vu_r^\star(\vx)=(G^\top G+rI)^{-1}G^\top\vx,$$
that minimizes
$\|G\vu-H\vx\|_2^2 + r\|\vu\|_2^2$,
and the ideal least squares solution, namely, $\vu^\star(\vx) = G^\dagger H\vx.$

\par The regularization parameters $\mu$ 
for the $\elll$ optimization
and $r$ for the $\ell^2$ optimization
are empirically chosen such that
the $\ell^2$ norm of the state $\{\vx(k)\}_{k=0}^{99}$
is minimized
($r=4.1042$ and $\mu=10.7167$).
For the $\ell^0$ optimization,
we choose the weighting matrix $Q$ in \eqref{eq:opt0} as $Q=I$,
and choose the matrix $W$ according to the procedure shown in Theorem \ref{thm:stability_l0}
with
$\E=\frac{2}{3}(1-\rho)P/c<(1-\rho)P/c$.

\begin{figure}[t!]
\centering
 \includegraphics[width=0.9\linewidth]{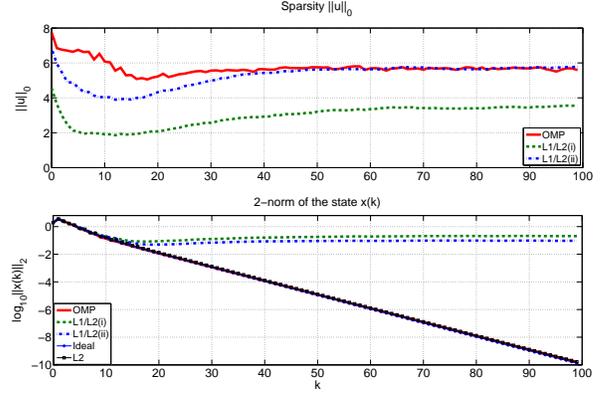}
\caption{
Averaged sparsity $\|\vu(\vx(k))\|_0$ of the control vectors (top) and averaged 2-norm
of the state $\vx(k)$ (bottom).
}
\label{fig:results}
\end{figure}

\par With  these parameters, we run 500 simulations
with randomly generated  packet-dropouts as described above,
and with initial vector $\vx_0$
in which each element is independently sampled
from the normal distribution with mean 0
and variance 1. 
The top figure in Fig.~\ref{fig:results} shows the averaged sparsity of the obtained
control vectors.
The $\elll$ optimization with $\mu=10.7167$,
labeled by L1/L2 (i),
always produces
sparser control vectors than  OMP.
Clearly, this property depends on how   the regularization parameter
$\mu>0$ is chosen.
In fact, if we choose the smaller value,  $\mu=3.3$, then the sparsity
approximates that of
OMP, as the curve labeled by L1/L2 (ii) in Fig.~\ref{fig:results}.
On the other hand, if we use a sufficiently large $\mu>0$,
then the control vector becomes $\vz$.
This is indeed the sparsest control, but leads
 to very poor control performance:
the state diverges until the control vector becomes
nonzero (in accordance with the stability results established in
Section~\ref{sec:stability}). 
The bottom figure in Fig.~\ref{fig:results}
shows the averaged 2-norm of the state $\vx(k)$ as a function of $k$ for all 5
designs.
\par 
We see that, with exception of the $\elll$ optimization based PPC,  the NCSs are all
asymptotically stable. (Our simulation results even suggest
exponential stability.) In contrast, if the  $\elll$
optimization of\cite{NagQue11} is used, then only practical stability is observed.
Note that the $\elll$ optimization with $\mu=3.3$ has almost the same sparsity
as OMP, but the response does not show asymptotic stability while OMP does.
The simulation  results are consistent with
Theorem~\ref{thm:stability_l1l2} and Theorem \ref{thm:stability_l0}.

\section{Conclusions}
\label{sec:conclusion}
We have studied packetized
predictive control formulations  with sparsity-promoting cost
functions (i.e., $\elll$ and $\ell^2$-constrained $\ell^0$
optimizations) for 
   networked control systems with  packet dropouts.
We have established sufficient conditions for practical stability of
$\elll$-optimal PPC and for asymptotic stability of $\ell^2$-constrained
$\ell^0$-optimal PPC,
when the number of successive packet dropouts is bounded.
Simulation results indicate that the proposed controllers provide, 
not only stabilizing but also sparse control packets.

Future work may include   obtaining analytical bounds
on the sparsity of solutions. 
It is also of interest to apply the proposed control methods to constrained
nonlinear plant models with disturbances, and to channels with bit-rate
limitations and unbounded packet-dropouts. We 
foresee that this will require extending results in\cite{QueNes12,QueOstNes11}
and also the development of fast  algorithms to solve the
associated optimization problems.

\section*{Acknowledgments}
The authors wish to thank the Associate Editor and  the anonymous 
reviewers for valuable comments which have helped to improve the 
quality of this note.

\bibliographystyle{IEEEtran}
\bibliography{IEEEabrv,ref}

\end{document}